# Gate tunable dark trions in monolayer WSe$_2$

Erfu Liu[1], Jeremiah van Baren[1], Zhengguang Lu[2,3], Mashael M. Altaiary[1], Takashi Taniguchi[4], Kenji Watanabe[4], Dmitry Smirnov[2], Chun Hung Lui[1*]

[1] Department of Physics and Astronomy, University of California, Riverside, CA 92521, USA.
[2] National High Magnetic Field Laboratory, Tallahassee, FL 32310, USA
[3] Department of Physics, Florida State University, Tallahassee, FL 32310, USA
[4] National Institute for Materials Science, Tsukuba, Ibaraki 305-004, Japan
[*] Corresponding author. Email: joshua.lui@ucr.edu

Abstract: We report the observation and gate manipulation of intrinsic dark trions in monolayer WSe$_2$. By using ultraclean WSe$_2$ devices encapsulated by boron nitride, we directly resolve the weak photoluminescence of dark trions. The dark trions can be tuned continuously between negative and positive charged trions with electrostatic gating. We also reveal their spin triplet configuration and distinct valley optical emission by their characteristic Zeeman splitting under magnetic field. The dark trions exhibit large binding energy (14 – 16 meV). Their lifetime (~1.3 ns) is two orders of magnitude longer than the bright trion lifetime (~10 ps) and can be tuned between 0.4 to 1.3 ns by electrostatic gating. Such robust, optically detectable, and gate tunable dark trions provide a new path to realize electrically controllable trion transport in two-dimensional materials.

Monolayer transition metal dichalcogenides (TMDs), such as $MoS_2$ and $WSe_2$, are remarkable two-dimensional (2D) semiconductors with strong Coulomb interactions [1, 2]. Their optical properties are dominated by tightly bound electron-hole pairs (excitons) at two time-reversal valleys (K, K') in the momentum space [3]. The strong spin-orbit coupling splits both the conduction and valence bands into two subbands with opposite spins [Fig. 1] [4-7]. The spin configuration governs an exciton's optical properties. If the electron and hole come from bands with the same electron spin, their recombination can efficiently emit light. These bright excitons have short lifetime (<10 ps) [8], in-plane dipole moment and out-of-plane light emission [Fig. 1] [9]. But if the electron and hole come from bands with opposite electron spins, the spin mismatch strongly suppresses their radiative recombination. They form dark excitons with long lifetime (>100 ps), out-of-plane dipole moment, and in-plane light emission [Fig. 1] [10-18]. Compared to bright excitons, the long-lived dark excitons are much better candidates for the studies of exciton transport and Bose-Einstein condensate [19-21], but their optical inactivity poses a significant challenge for experiment.

Monolayer $WSe_2$ is an exceptional material to explore dark excitons. Unlike other semiconductors (*e.g.* $MoSe_2$) with bright excitons in the lowest energy level, monolayer $WSe_2$ hosts dark excitons well below the bright exciton level [Fig. 1] [6, 13, 22]. The dark excitons can thus accumulate a sufficiently large population to achieve observable light emission, as reported by prior research [16, 23-26]. Moreover, as the dark excitons lie at the lowest energy level, they play a crucial role in the carrier dynamics of monolayer $WSe_2$. It is therefore important to understand their diverse properties.

Similar to bright excitons/trions, dark excitons can capture an extra charge to form dark trions [Fig. 1]. The dark trions are fascinating entities for excitonic transport, because their finite net charge, together with their long lifetime, would enable effective control of exciton dynamics by electric field. However, detection and manipulation of dark trions are challenging due to their optical inactivity. Recent research has revealed evidence of dark trions in monolayer $WSe_2$ under strong in-plane magnetic field [14] and plasmonic environment [15]. But these indirect detection methods are difficult to apply for practical applications. They also inevitably distort the trion properties (e.g. twisting of spins by in-plane magnetic field). It is preferable to directly probe the intrinsic properties of dark trions. However, due to many difficulties in device quality and experiment, direct probe of gate-tunable intrinsic dark trions has not been demonstrated in 2D materials yet.

In this letter, we carry out a comprehensive investigation of the optical emission from the intrinsic dark trions in monolayer $WSe_2$. Our experiment is made possible with ultraclean monolayer $WSe_2$ devices encapsulated by boron nitride (BN). The supreme quality of our devices allow us to observe both the positive and negative charged dark trions under continuous electrostatic gating. We reveal the spin triplet configuration and distinct valley optical emission of dark trions by their characteristic Zeeman splitting under strong out-of-plane magnetic field. The g-factors (~ –9) of dark trions with spin triplet are about twice the g-factors (~ –4) of bright trions with spin singlet. Compared to the bright trions with binding energies 21 – 35 meV, the dark trions have smaller, but still sizable, binding energies (14 – 16 meV). Notably, their lifetime (1.3 ns) is two orders of magnitude longer than the bright trion lifetime (~10 ps) and can be tuned continuously between 0.4 and 1.3 ns by electrostatic gating. Such robust, optically detectable, and gate tunable dark trions hold promises for realizing field-controlled trion transport in 2D materials.

In the experiment, we measure the photoluminescence (PL) from BN-encapsulated monolayer $WSe_2$ devices with 532-nm continuous laser excitation at temperature T ~ 4 K. The details of device fabrication and experimental methods are presented in the Supplemental Material [27]. We applied very low incident laser power (16 μW) to suppress the biexciton PL, which is known to be significant in monolayer $WSe_2$ [23-27]. We find that, in our ultraclean samples, dark trions can emit weak but noticeable PL. Although such PL propagates in the in-plane direction, we can partially capture it with a wide-angle microscope objective (N.A. = 0.67) in the conventional out-of-plane detection geometry [16]. This detection scheme is much simpler than the previous indirect detection methods, which require either in-plane magnetic field, plasmonic coupling, or near-field tip enhancement [11, 12, 14, 15].

Fig. 2(a) displays a gate-dependent PL map of monolayer $WSe_2$. Fig. 2(b) displays three representative PL spectra at gate voltages $V_g$ = 0, –0.5, and –1.5 V, which correspond to the electron side, charge neutrality point, and hole side, respectively. We observe the bright A exciton ($A^0$) at 1.712 meV at the charge neutrality point. As we tune $WSe_2$ to the electron and hole sides, we observe one positive ($A^+$) and two negative ($A_1^-$, $A_2^-$) bright trions; their binding energies are 21, 29 and 35 meV, respectively [Fig. 2(a-c)]. These energy values are consistent with prior reports [32, 33].

At the charge neutrality point, monolayer $WSe_2$ exhibits a weak PL peak at 1.672 eV, 41 meV below the bright exciton ($A^0$) [Fig. 2(a-c)]. This is the dark exciton ($D^0$) according to prior studies [12, 14-16, 23-26]. As we tune monolayer $WSe_2$ to the electron and hole sides, the $D^0$ peak subsides and two new peaks emerge at 16 and 14 meV below the $D^0$ peak. We tentatively denote them as $D^-$ and $D^+$, respectively (other peaks are discussed in the Supplemental Material [27]). Both the $D^-$ and $D^+$ intensity increases linearly with the excitation laser power; they are thus not associated with biexcitons with quadratic power dependence [27]. Below, we will prove that the $D^-$ and $D^+$ features come from the dark trions by their distinctive characteristics.

First, we have compared the gate-dependent PL energy, intensity and line width of the $D^0$, $D^-$, $D^+$ peaks with those of the bright exciton/trions ($A^0$, $A^+$, $A_1^-$, $A_2^-$) [(Fig. 2(c-e)]. These two sets of peaks exhibit parallel gate-dependent behavior. This suggests that the $D^-$ and $D^+$ peaks correspond to the negative and positive dark trions, respectively. Moreover, the energy separation (14 – 16 meV) between the $D^-$, $D^+$ peaks and the $D^0$ peak matches the reported binding energy of dark trions in monolayer WSe$_2$ [14, 27]).

Second, when we decrease the temperature, the bright trion PL drops continuously, but the dark trion PL grows stronger (see Fig. S8 in the Supplementary Materials [27]). Such contrasting behavior reflects their different energy levels and population distribution. The bright trions lie at higher energy and their population drops with decreasing temperature, but the dark trions lie at the lowest energy and their population increases at low temperature [Fig. 1] [13].

Third, the dark trions have a distinct spin configuration from that of bright trions. While the bright trions involve a spin-singlet exciton, the dark trions involve a spin-triplet exciton (see Fig. 1; the hole has opposite spin of the electron in the valence band) [18]. As we will show below, these spin configurations can be identified by the Zeeman effect.

Fourth, the dark trions follow different optical selection rules from those of bright trions. According to prior research, the bright excitons at the K (K') valley emit light with right-handed (left-handed) circular polarization in the out-of-plane direction [5, 34, 35], but the dark excitons at both K and K' valleys emit light with vertical linear polarization in the in-plane direction [Fig. 1] [16]. Their associated bright/dark trions follow the same optical selection rules, because the third charge only affects weakly the recombination of the component exciton.

We can confirm the third and fourth characteristics of dark trions by measuring their PL under out-of-plane magnetic field. Prior research has demonstrated that out-of-plane magnetic field can lift the valley degeneracy in monolayer TMDs [30, 36-38]. Due to the opposite spin and orbital configurations of the K and K' valleys, the magnetic field will enlarge the band gap at one valley but diminish the band gap at the other valley [Fig. 3(a)]. This will induce the valley Zeeman splitting $\Delta E = g\mu_B B$, where g is the effective g-factor and $\mu_B$ is the Bohr magneton. For the bright exciton/trions, their Zeeman splitting arises mainly from the orbital angular momentum of the valance band, with a total g-factor close to –4 [30, 36-38]. But for the dark exciton/trions, their associated spin triplet will contribute an additional g-factor of –4. This will make their total g-factor about twice of that of bright trions [18, 31].

To further probe the distinct optical selection rules of dark and bright trions, we adopt a special measurement geometry – we excite monolayer WSe$_2$ with unpolarized light but collect only the PL with left circular polarization. For the bright exciton/trions with circularly polarized PL, our measurement only detects emission from the K' valley. But for the dark exciton/trions with linearly polarized PL, our measurement detects emission from both valleys.

We have studied monolayer WSe$_2$ under out-of-plane magnetic field from B = –31 to 31 T [Fig. 3]. Fig. 3(b) displays the gate-dependent PL map of monolayer WSe$_2$ at B = –

10 T. The spectra of the bright exciton/trions remain largely unchanged except for a Zeeman energy shift – we detect light from only one valley as expected from bright exciton/trions. The extracted g-factors (–4.1 to –4.4) of the bright exciton/trions are consistent with their spin singlet configuration as well as the values in the literature [Fig. 3 (c-d)] [30, 36-38]. In contrast, the $D^0$, $D^-$, $D^+$ peaks each split into two peaks with the same energy separation – we detect light emission from both valleys as expected from dark exciton/trions [Fig. 3(a-c)]. Moreover, their extracted g-factors (–9.1 to –9.5) are about twice of the g-factors of bright exciton/trions [Fig. 3(d)] [27]. Such large g-factors are consistent with the spin triplet of dark exciton/trions [18, 31]. Our experiment therefore confirms both the spin triplet configuration and the distinct valley optical selection rules for the dark trions in monolayer WSe$_2$.

Finally, dark trions are expected to live much longer than the bright trions, because their spin triplet configuration suppresses the radiative decay [14, 18]. The long lifetime of dark trions can be inferred from the line widths of the $D^-$ and $D^+$ peaks (FWHM ~ 2.5 meV), which are much narrower than the bright trion line widths (FWHM = 3 – 9 meV) [Fig. 2(e)]. To directly probe their lifetime, we have measured the time-resolved PL intensity of trions by the time-correlated single photon counting method [27]. Fig. 4(a) displays the time-resolved PL trace of the positive bright trion ($A^+$) and dark trion ($D^+$). The bright trion dynamics exhibits two exponential decay components. The fast and dominant component follows closely the instrument response function (IRF) of our setup. We extract a lifetime of ~10 ps after deconvolution with the IRF. The slow component contributes little because it is an order of magnitude weaker than the fast component [inset of Fig. (4a)] [39-41]. In contrast, the dark trion (at $V_g$ = –0.8 V) shows a single exponential decay with a lifetime of 1270 ps, which is two orders of magnitude longer than the bright trion lifetime (~10 ps).

Notably, the dark trion dynamics depends strongly on the charge density and correlates with the PL intensity and line width. Fig. 4(b) displays the dark trion lifetime, PL intensity and line width as a function of gate voltage ($V_g$). When we tune $V_g$ from 0 V to –1 V in the hole side, the dark trion lifetime increases slightly from 1 ns to 1.3 ns; the PL intensity rises; the line width remain almost constant. As we tune $V_g$ further to –4.4 V, the dark trion lifetime decreases from 1.3 ns to 370 ps; the PL intensity decreases; the line width increases correspondingly [27]. These observations can be qualitatively understood from the interactions between trions and free carriers. When carriers are first injected into the sample, they facilitate the trion formation and enhance the trion PL. But as the carrier density continues to increase, the free carriers will scatter frequently with the trions. Such scattering will shorten the trion lifetime, suppress the trion PL, and broaden the PL peak.

In conclusion, we have observed and identified the dark trions in monolayer WSe$_2$ by their characteristic gate and temperature dependence, Zeeman splitting, valley optical selection rules, and long lifetime. Prior research has shown that trions can behave like free charge carriers with controlled motion under electric field [42, 43]. Given their charge, spin, valley and layer degree of freedom, TMD trions are attractive carriers for quantum information technology. But trion transport has so far not been achieved in 2D materials, primarily due to the short lifetime of bright trions. Our observed long-lived dark trions offer an effective solution. In a rough estimation of the trion drift distance, we may assume the trion mobility ($\mu$) to be 1/10 of the electron/hole mobility, considering the trion's triple mass and higher scattering rate. The trion mobility can reach $\mu$ = 400 cm$^2$V$^{-1}$s$^{-1}$ from the highest carrier mobility (4,000 cm$^2$ V$^{-1}$ s$^{-1}$) reported in monolayer WSe$_2$ [44]. By using an electric field $E = 0.5$ V/μm and our observed lifetime $\tau = 1\ ns$, the drift distance of dark trions under electric field can reach $l_d = \mu E \tau = 20\ \mu m$. This value exceeds the typical length scale of 2D material devices. We should be able to drive the dark trions across the device and image their motion by their luminescence. The long-lived, optically detectable and gate tunable dark trions therefore provide a new path to investigate field-controlled trion transport.

**Acknowledgements**
We thank N. M. Gabor for discussion. Z.L. and D.S. acknowledge support from the US Department of Energy (DE-FG02-07ER46451) for magneto-photoluminescence measurements performed at the National High Magnetic Field Laboratory, which is supported by the NSF Cooperative Agreement no. DMR-1644779 and the State of Florida. K.W. and T.T. acknowledge support from the Elemental Strategy Initiative conducted by the MEXT, Japan and the CREST (JPMJCR15F3), JST.

**Figure 1**

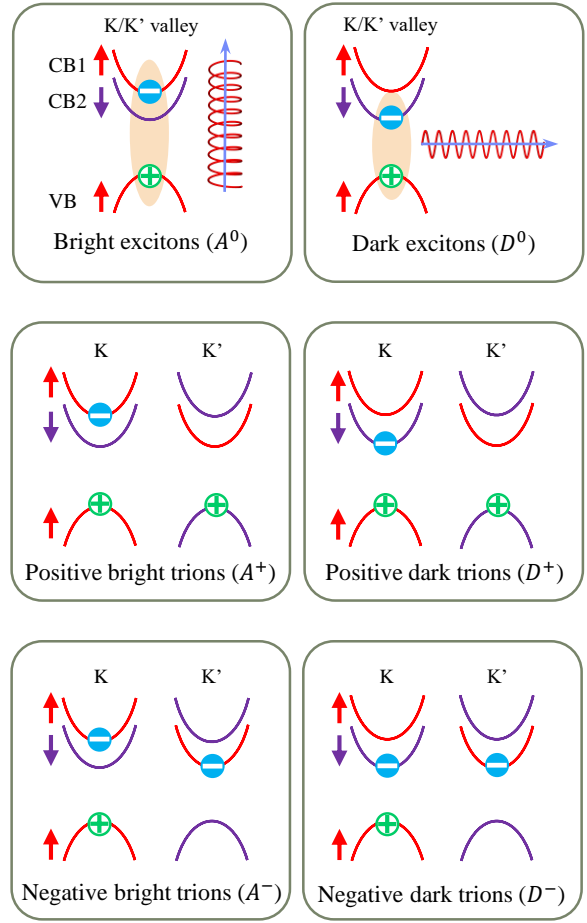

FIG. 1. Schematic configurations of bright/dark excitons/trions in the conduction bands (CB1, CB2) and valance band (VB) at the K and K' valleys in monolayer WSe$_2$. The arrows and colors denote the electron spin direction. The bright excitons/trions emit circularly polarized light in the out-of-plane direction; the dark excitons/trions emit vertically polarized light in the in-plane direction. In this simple picture, we ignore possible contributions from the Q valleys to the exciton/trion properties.

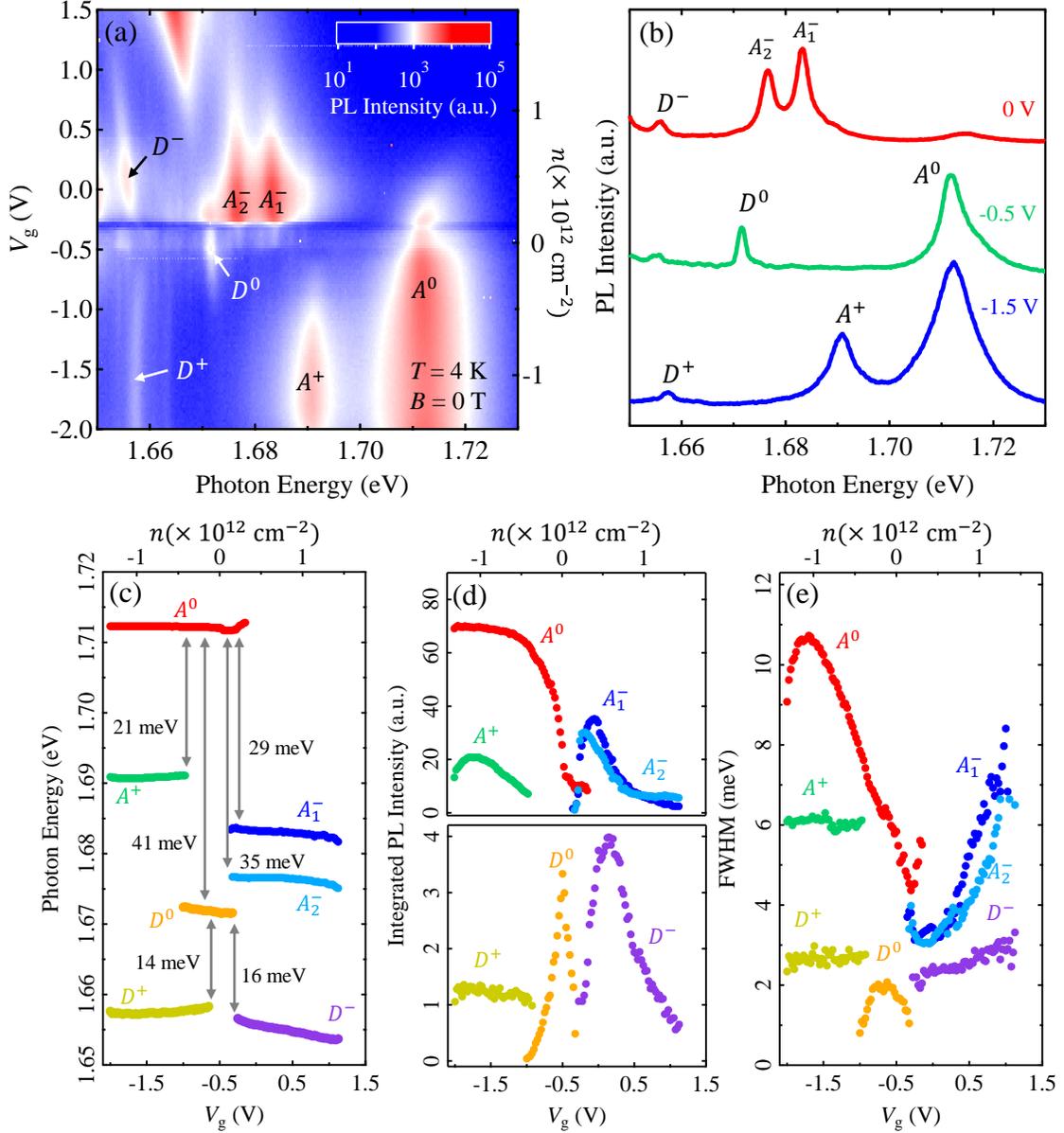

FIG. 2. Observation of dark trions in monolayer WSe$_2$. (a) The gate-dependent photoluminescence (PL) map of monolayer WSe$_2$ (Device 1) at T = 4 K and B = 0 T under 532-nm continuous laser excitation. (b) The cross-cut spectra at gate voltages $V_g$ = 0, –0.5, –1.5 V. (c) The extracted PL energy, (d-e) PL integrated intensity, (e) PL full width at half maximum (FWHM) of dark and bright excitons/trions as a function of gate voltage (bottom axis) and charge density (top axis).

**Figure 3**

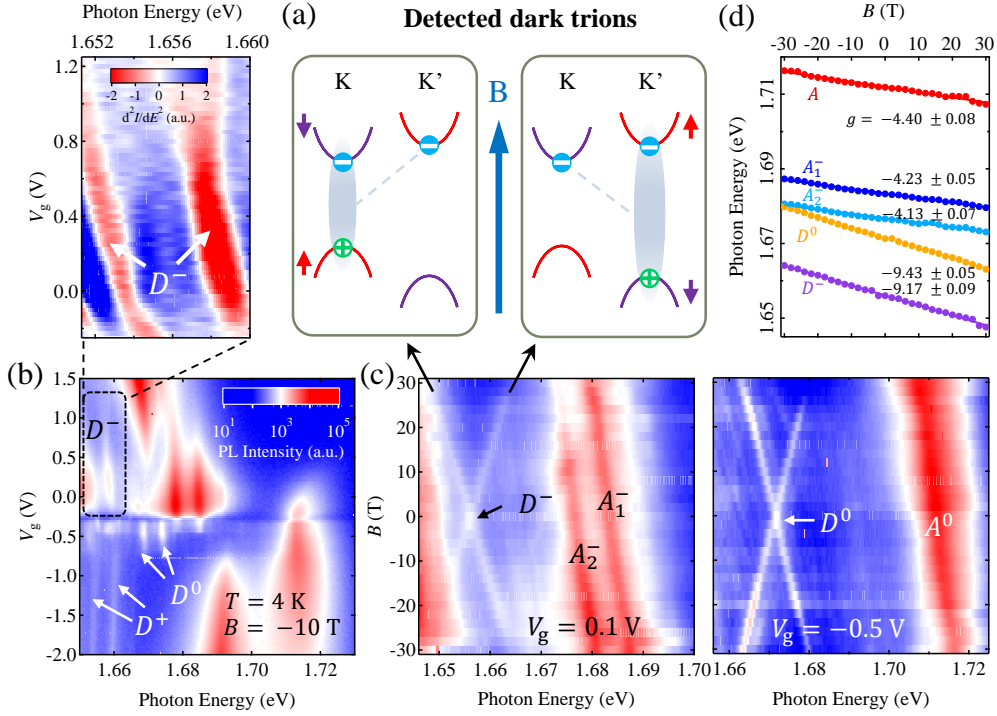

FIG. 3. (a) Configurations of the negative dark trions detected in our magneto-PL experiment. The two valleys have different energy gaps under magnetic field. (b) The PL map of monolayer WSe$_2$ (Device 1) at $T = 4$ K and $B = -10$ T. We excite monolayer WSe$_2$ with unpolarized light and collect only the PL with left helicity. The dark exciton/trions are split into two peaks, whereas the bright exciton/trions are not. We take the second energy derivative ($d^2I/dE^2$) of the PL spectra inside the dashed box to highlight the splitting of $D^-$ trion. (c) The $B$-dependent PL map at the electron side ($V_g = 0.1$ V) and charge neutrality point ($V_g = -0.5$ V). Panels b and c share the same color scale bar. (d) The Zeeman energy shifts of the excitons/trions. The g-factors are extracted by linear fits.

**Figure 4**

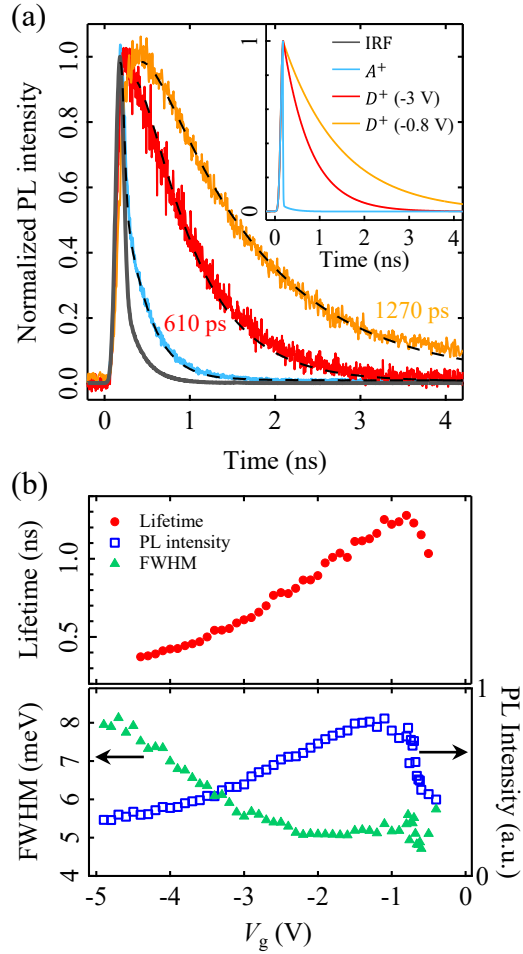

FIG. 4. (a) Time-resolved PL for the positive bright trion ($A^+$) and dark trion ($D^+$) in monolayer $WSe_2$ (Device 2). The black solid line is the instrument response function (IRF). We fit the $D^+$ and $A^+$ data, respectively, with single-exponential and biexponential functions convolved with the IRF (dashed lines). The inset shows the deconvolved fits. The $D^+$ lifetime is 1270 ps and 610 ps at gate voltages $V_g$ = –0.8 V and –3 V, respectively. (b) The lifetime, PL intensity and full width at half maximum (FWHM) of $D^+$ trion as a function of gate voltage.

# Supplementary Material of
# "Gate tunable dark trions in monolayer WSe$_2$"

Erfu Liu[1], Jeremiah van Baren[1], Zhengguang Lu[2,3], Mashael M. Altaiary[1], Takashi Taniguchi[4], Kenji Watanabe[4], Dmitry Smirnov[2], Chun Hung Lui[1]*

[1] Department of Physics and Astronomy, University of California, Riverside, CA 92521, USA.

[2] National High Magnetic Field Laboratory, Tallahassee, FL 32310, USA

[3] Department of Physics, Florida State University, Tallahassee, FL 32310, USA

[4] National Institute for Materials Science, Tsukuba, Ibaraki 305-004, Japan

* Corresponding author. Email: joshua.lui@ucr.edu

## 1. Device fabrication

The BN-encapsulated WSe$_2$ devices are fabricated by mechanical co-lamination of two-dimensional (2D) crystals. We use WSe$_2$ bulk crystals from HQ Graphene Inc. We first exfoliate monolayers WSe$_2$, multi-layer graphene and thin BN flakes from their bulk crystals onto the Si/SiO$_2$ substrates. Afterward, a polycarbonate-based dry-transfer technique is applied to stack the different 2D crystals together. We use a stamp to first pick up a BN flake, and sequentially pick up multi-layer graphene (as the electrodes), a WSe$_2$ monolayer, a BN thin layer (as the bottom gate dielectric), and a graphene multi-layer (as the back gate). This method ensures that the WSe$_2$ layer doesn't contact the polymer during the whole fabrication process, so as to reduce the contaminants and bubbles at the interface. Afterward, standard electron beam lithography is applied to pattern and deposit the gold contacts (100 nm thickness). Finally, the devices are annealed at 300 °C for 3 hours in an argon environment. Fig. S1 displays the schematic of our devices and the optical image of a representative device. For Device 1 used in Fig. 1–3 of the main paper, the bottom BN gate dielectric layer is 24.5 nm in thickness. For Device 2 used in Fig. 4 and Fig. S6-7, the bottom BN thickness is about 35 nm. We can deduce the capacity and gating charge density of the back gate when we know the BN thickness.

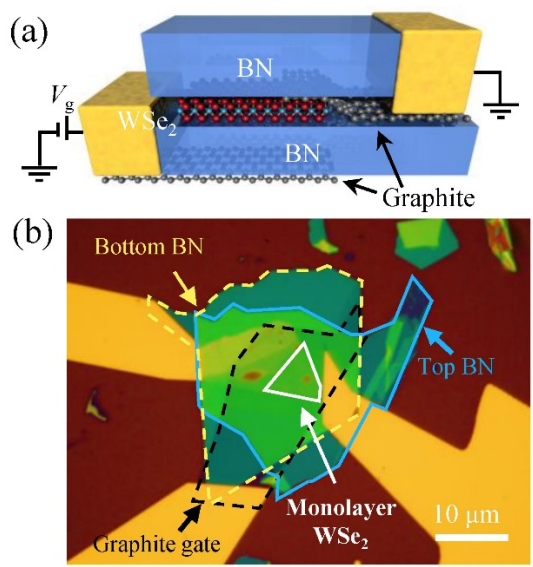

**Fig. S1** (a) The schematic of BN-encapsulated monolayer WSe$_2$ device. (b) The optical image of a representative device.

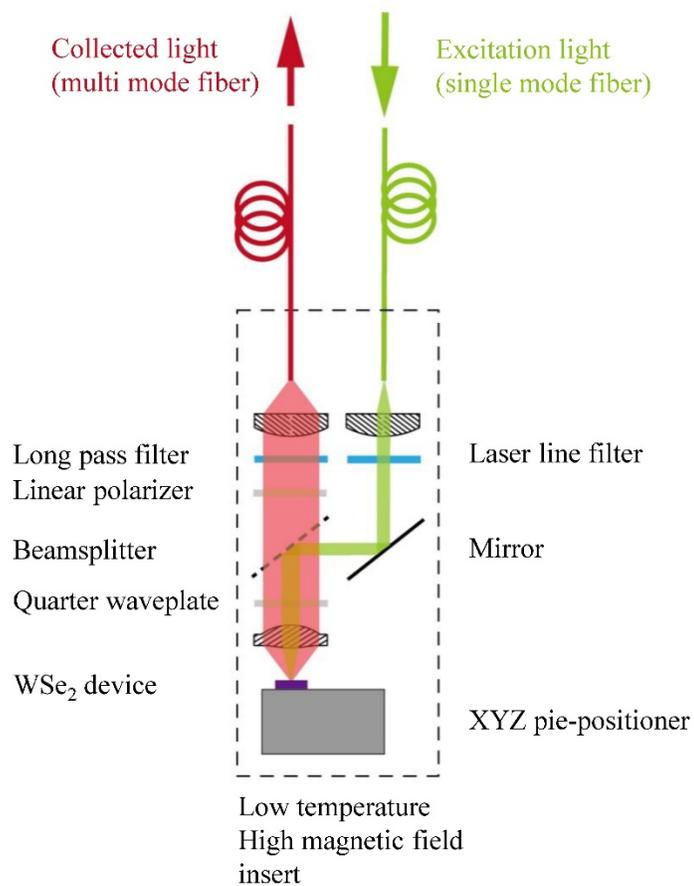

**Fig. S2.** Fiber-based probe setup for PL measurement under high magnetic field.

## 2. Experimental methods

We performed the magneto-optical experiment in the National High Magnetic Field Lab (NHMFL) with 31 T DC magnet by using a fiber-based probe on Device 1. Fig. S2 displays the schematic of the experimental setup. We use a 532-nm continuous laser as the excitation light source. The laser beam is transmitted through a single-mode optical fiber and focused by a lens (NA = 0.67) onto the sample. The sample is mounted on a three-dimensional piezoelectric translation stage. The photoluminescence (PL) is collected through a 50/50 beam splitter into a multimode optical fiber, and subsequently measured by a spectrometer with a CCD camera (Princeton Instruments, IsoPlane 320). A quarter wave plate is used to select the left-handed circularly polarized component of the PL signal.

The time-resolved PL experiment was conducted by the time-correlated single photon counting technique in our laboratory at UC Riverside. The excitation laser pulses come from an oscillator (Light Conversion Inc., Pharos) with 1030 nm output wavelength, 80 MHz repetition rate, and 90 fs pulse duration. We use the second harmonic (515 nm wavelength) of the laser to excite our samples, which are mounted in a Montana cryostat. The PL signal is first spectrally dispersed and filtered by a grating monochromator. We only select the PL from the dark or bright excitons/trions in our experiment. We detect the PL with an avalanche photodiode (PicoQuant, PDM) and measure its temporal trace by time-correlated single photon counting (PicoQuant, PicoHarp 300). The instrument response function (IRF) of our time-resolved PL setup is determined from the response of our laser pulses. We obtain a time resolution of ~50 ps from the instrument response.

In the main paper, we have presented the PL lifetime of the positive dark trion ($D^+$). The measurement of the $D^+$ lifetime is reliable because this feature is isolated from other PL peaks. But the negative dark trion ($D^-$) at 1.655 eV is somewhat connected to a few bright and broad peaks at 1.62 – 1.65 eV (see the PL image at Fig. S5a). These low-energy bright peaks contribute a large background to the $D^-$ PL signals and make the lifetime measurement unreliable. Therefore, we only discuss the lifetime of the positive dark trion in the main paper. We expect similar lifetime and gate dependence for the negative dark trion.

## 3. Power dependence of photoluminescence from dark excitons/trions

We have measured the PL from bright exciton and dark exciton/trions on Device 1 at different excitation laser power (Fig. S3). The measurements were performed in an optical microscopic station with no magnetic field. The station consists of a 532-nm continuous laser (Torus 532), a commercial microscope (Nikon Eclipse T*i*-U), a helium-cooled cryostat (Janis ST-500), and a high-resolution spectrometer (Princeton Instruments; Isoplane 320). The excitons and trions all show approximately linear power dependence. The power-law fits $I = P^\alpha$ give the exponent of $\alpha$ = 0.88, 1.14, 1.04 and 0.90 for the $A^0$, $D^+$, $D^0$ and $D^-$ peaks, respectively.

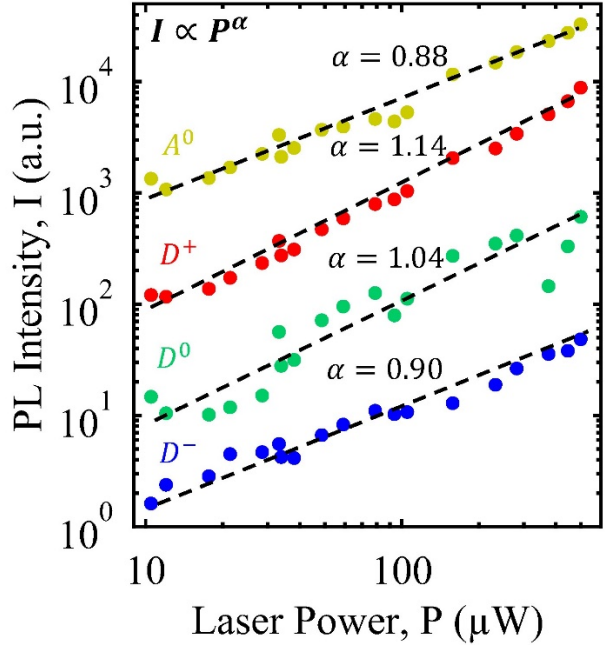

**Fig. S3.** PL intensity as a function of excitation laser power for the bright exciton ($A^0$), dark exciton ($D^0$), positive and negative dark trion ($D^+$, $D^-$). The dash lines are the power-law fits $I = P^\alpha$.

### 4. Gate-dependent photoluminescence mapping under various laser power

We have measured the gate-dependent PL maps on Device 1 with five different excitation laser power, $P$ = 16, 50, 100, 200, 500 μW (Fig. S4). We can clearly observe the dark exciton/trions at low excitation power ($P$ = 16 μW). But when the laser power increases, the PL signals of biexciton ($AD^0$) and charged biexciton ($AD^-$) become bright and obscure the signals of dark exciton/trions [1-4]. Therefore, we have to use low laser power to study the dark trions. The exciton and trions typically exhibit linear dependence on the excitation laser, whereas the biexcitons exhibit quadratic dependence. We can therefore readily identify the biexciton and charged biexciton from their quadratic dependence on the excitation laser intensity (Fig. S4f)

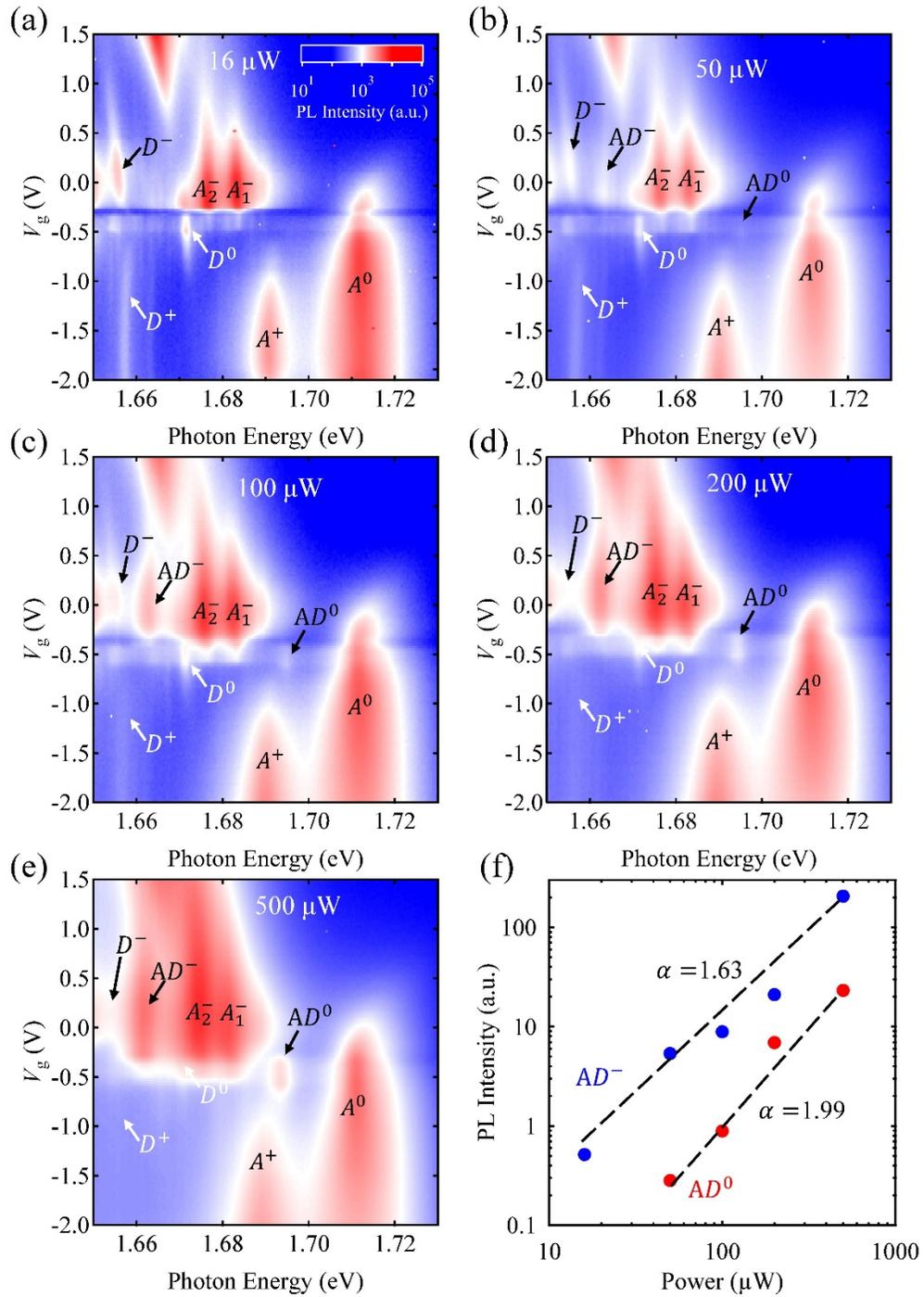

**Fig. S4.** Gate-dependent PL maps under different excitation laser power: (a) 16 μW, (b) 50μW, (c) 100μW, (d) 200μW, (e) 500 μW. (f) PL intensity (*I*) as a function of laser power (*P*) for the neutral biexciton ($AD^0$) and charged biexciton ($AD^-$). They both show superlinear power dependence with exponent *α* = 1.99 and 1.63, respectively, in the power-law fits $I = P^\alpha$ (dashed lines).

## 5. Other emission features in the photoluminescence map

Besides the excitons, trions and biexcitons, our PL maps also reveal several peaks with unknown origin. Fig. S5 displays a PL image in a wider photon energy range. We observe several new features at energies of 1.62 – 1.65 eV below the $D^-$ and $D^+$ features. These new features exhibit different g-factors from -4 to -12, as seen in the magnetic-field-dependent PL map (Fig. 5b-c). They may arise from the momentum-indirect excitons/trions between the K valley and Q valley [2, 5]. In addition, a new feature ($A_3^-$) appears at 1.66 – 1.67 eV at high gate voltage ($V_g > 1$ V). It may arise from the trions formed between an exciton and an electron in the high-lying conduction band. Further experiments are needed to determine the origins of these new features.

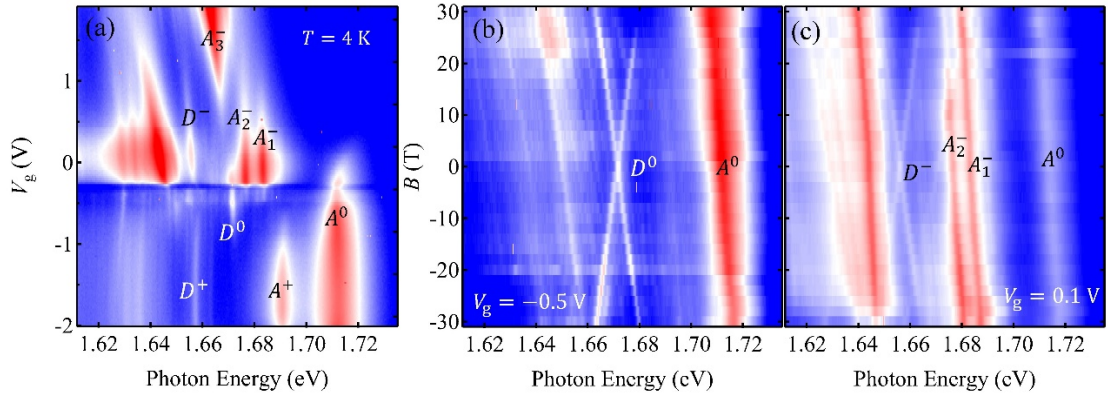

**Fig. S5.** PL maps in a wider photon energy range. (a) Gate-dependent PL map. (b-c) Magnetic-field-dependent PL maps at gate voltages $V_g = -0.5$ V and 0.1 V.

## 6. Summary of PL and binding energy of excitons, trions and biexcitons

As shown in the PL images in Fig. S4, we have observed a panoply of excitonic species in monolayer WSe2, including the bright exciton and trions ($A^0$, $A_1^-$, $A_2^-$, $A^+$), dark exciton and trions ($D^0$, $D^-$, $D^+$), biexciton ($AD^0$) and charged biexciton ($AD^-$). It would be useful to summarize the properties of these excitonic states. Table S1 below list the measured PL energy and binding energy of all these excitonic states in our experiment. Our values are comparable to those in the literature.

| Exciton states | PL energy (eV) | | Binding energy (meV) | |
| --- | --- | --- | --- | --- |
| | Our results | Prior results | Our results | Prior results |
| Bright exciton ($A^0$) | 1.7121 | 1.728 [1]<br>1.723 [2]<br>1.740 [3]<br>1.750 [6]<br>1.738 [7]<br>1.722 [8] | - | - |
| Dark exciton ($D^0$) | 1.6716 | 1.685 [1]<br>1.681 [2]<br>1.697 [3]<br>1.704 [6]<br>1.696 [7]<br>1.682 [8] | 40.5 | 43 [1]<br>42 [2]<br>43 [3]<br>46 [6]<br>42 [7]<br>40 [8] |
| Negative bright trion ($A_1^-$) | 1.6835 | 1.699 [1]<br>1.693 [2]<br>1.712 [3]<br>1.690 [8] | 28.6 | 29 [1]<br>30 [2]<br>28 [3]<br>32 [8] |
| Negative bright trion ($A_2^-$) | 1.6767 | 1.692 [1]<br>1.687 [2]<br>1.705 [3] | 35.4 | 36 [1]<br>36 [2]<br>35 [3] |
| Negative dark trion ($D^-$) | 1.6564 | 1.683 [6] | 15.2 | 21 [6] |
| Positive bright trion ($A^+$) | 1.6911 | 1.703 [1]<br>1.701 [2]<br>1.701 [3] | 21.0 | 25 [1]<br>22 [2]<br>39 [3] |
| Positive dark trion ($D^+$) | 1.6579 | - | 13.7 | - |
| Biexciton ($AD^0$) | 1.6933 | 1.711 [1]<br>1.703 [2]<br>1.723 [3] | 18.8 | 17 [1]<br>20 [2]<br>17 [3] |
| Charged biexciton ($AD^-$) | 1.6615 | 1.679 [1]<br>1.671 [2]<br>1.691 [3] | 50.6 | 49 [1]<br>52 [2]<br>49 [3] |

**Table S1.** Summary of our measured PL energy and binding energy of different excitonic species in monolayer WSe$_2$, in comparison with the values in the literature.

## 7. Photoluminescence measurements on monolayer WSe$_2$ Device 2

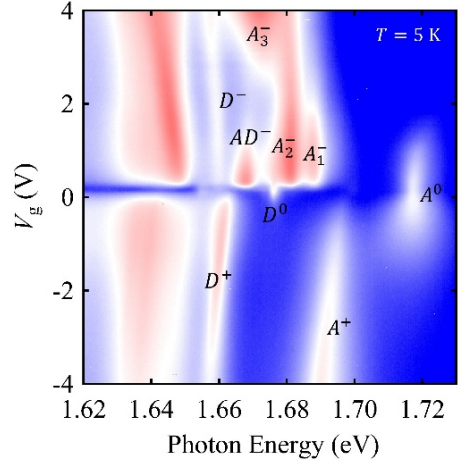

**Fig. S6.** Gate-dependent PL map on BN-encapsulated monolayer WSe$_2$ Device 2 at T = 5 K and B = 0 K. The incident laser power is 20 μW.

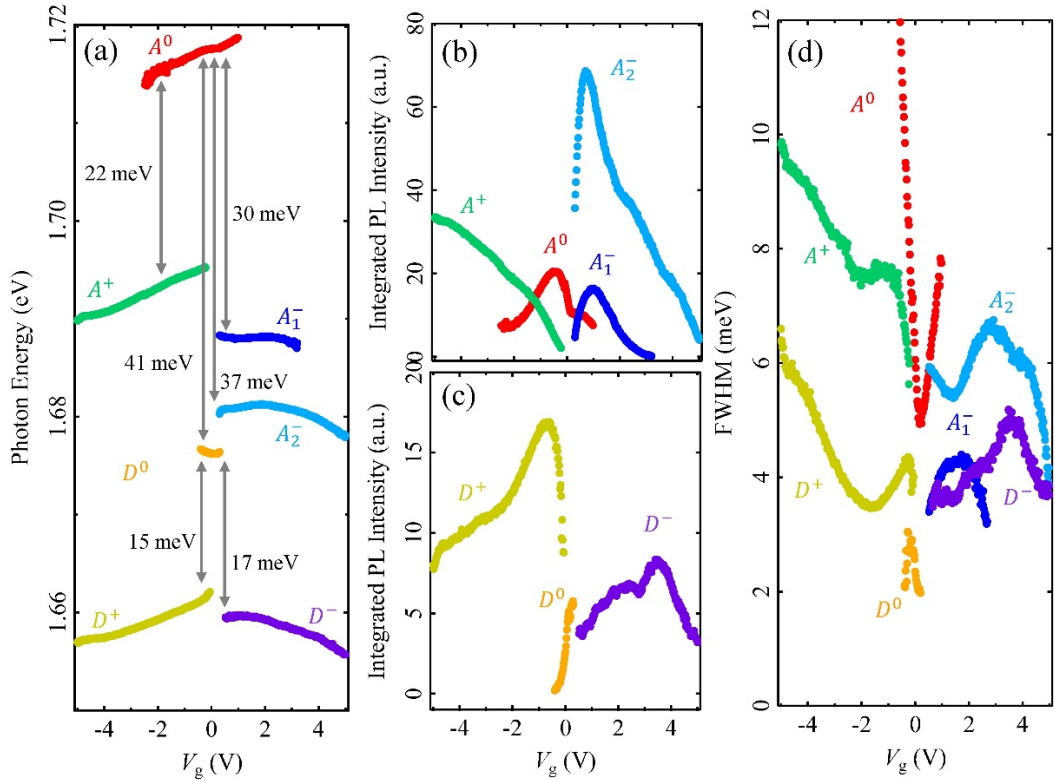

**Fig. S7.** The gate-dependent exciton/trion parameters in monolayer WSe$_2$ Device 2. (a) The PL energy, (b-c) integrated PL intensity, (d) PL full width at half maximum (FWHM) of the dark and bright excitons/trions as a function of gate voltage (bottom axis) and charge density (top axis). All parameters are extracted from the PL data in Fig. S6.

We have measured the PL on several BN-encapsulated monolayer WSe$_2$ devices. They show similar behavior. For instance, Fig. S6 displays the gate-dependent PL map of Device 2. Fig. S7 displays the PL energy, intensity, line width of the bright and dark exciton/trions in Device 2 as a function of gate voltage. The results are qualitatively similar to those of Device 1 (Fig. 2 of the main paper).

## 8. Temperature-dependent photoluminescence of bright and dark trions

We have measured the PL spectra of bright trion (A$^+$) and dark trion (D$^+$) at different temperatures from 5 to 75 K (Fig. S8). As the temperature decreases, the bright trion peak becomes weaker, but the dark trion peak grows stronger. Similar opposite temperature-dependent PL has been observed in bright and dark excitons in monolayer WSe$_2$ [9]. The result can be understood from the different energy levels of bright and dark trions. In monolayer WSe$_2$, the bright trions occupy an elevated energy level. Their population (and hence the PL intensity) decreases at low temperature, when the thermal energy is not sufficient to sustain the high-lying bright trions. In contrast, the dark trions occupy the lowest energy level. Their population accumulates and increases at low temperature, leading to stronger PL at low temperature.

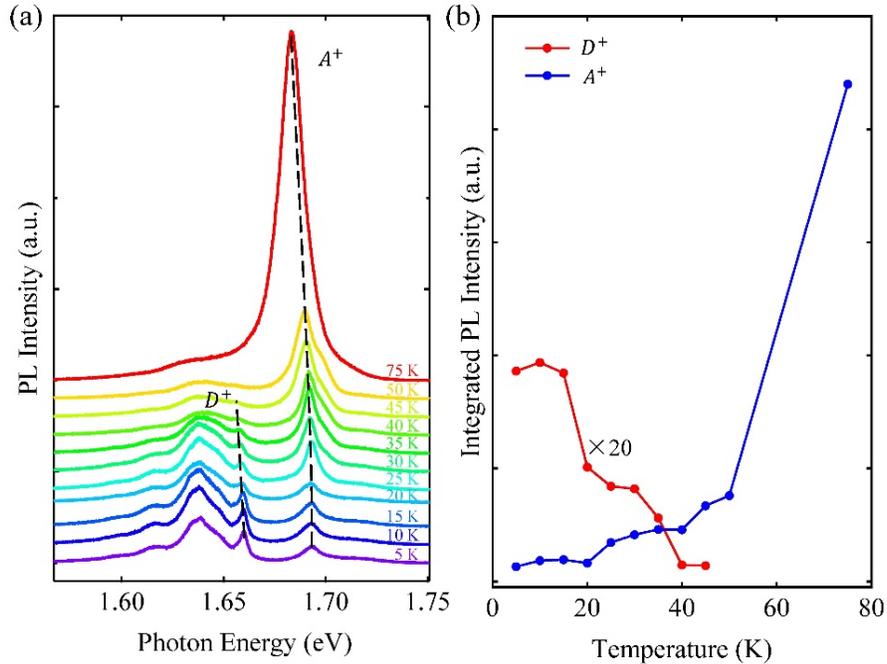

**Fig. S8.** Temperature-dependent PL of bright and dark trions in monolayer WSe$_2$. (a) PL spectra at different temperatures from 5 to 75 K. (b) The integrated PL intensity of dark trion (D$^+$) and bright trion (A$^+$) as a function of temperature. The gate voltage is $V_g = -2$ V.

## 9. Zeeman splitting of dark exciton and trions in monolayer WSe$_2$

The K and K' valleys in monolayer WSe$_2$ are connected by the time-reversal symmetry and hence energy-degenerate. Such a degeneracy can be lifted by breaking the time-reversal symmetry with the out-of-plane magnetic field, leading to the valley Zeeman splitting. For the bright exciton, the Zeeman splitting can be revealed by the energy splitting of light emission with opposite circular polarization [10-13]. The valley Zeeman splitting of bright exciton is $\sim -4\mu_B B$, where $\mu_B$ is the Bohr magneton and B is the magnetic field. Such a splitting comes mainly from the intracellular orbital magnetic moment (0 for the conduction band and $\pm 2\mu_B B$ for the valance band in different valleys). The contribution from the spin magnetic moment and spin-orbit coupling is small for the bright exciton with spin singlet configuration. For the dark exciton with spin triplet configuration, the spins contribute an additional Zeeman splitting of $-4\mu_B B$. The total Zeeman splitting of dark exciton is thus expected to be twice of that of bright exciton [1, 3, 4, 14, 15]. The Zeeman effect of the bright/dark trions is similar to the associated bright/dark exciton, because the properties of the emitted photon is determined mainly by the recombination of the exciton.

In Fig. 3 of the main paper, we only collect the left-handed circularly polarized PL light. This measurement geometry detects the bright exciton/trions only in one valley; we therefore need to use their energy in positive and negative magnetic field to determine their Zeeman splitting. But for the dark exciton/trions with linearly polarized in-plane emission, we detect their light emission from both valleys and hence directly visualize their Zeeman splitting.

Fig. S9 displays the PL spectra of monolayer WSe$_2$ under different magnetic fields at gate voltages $V_g$ = 0.5 V and 0.1 V. Fig. S10 displays the energy of the dark exciton and trion as a function of magnetic field. We can clearly see the evolution of the dark exciton and trion as a function of magnetic field and their valley Zeeman splitting. The data in Fig. S9 and S10 correspond to the data in Fig. 3c of the main paper.

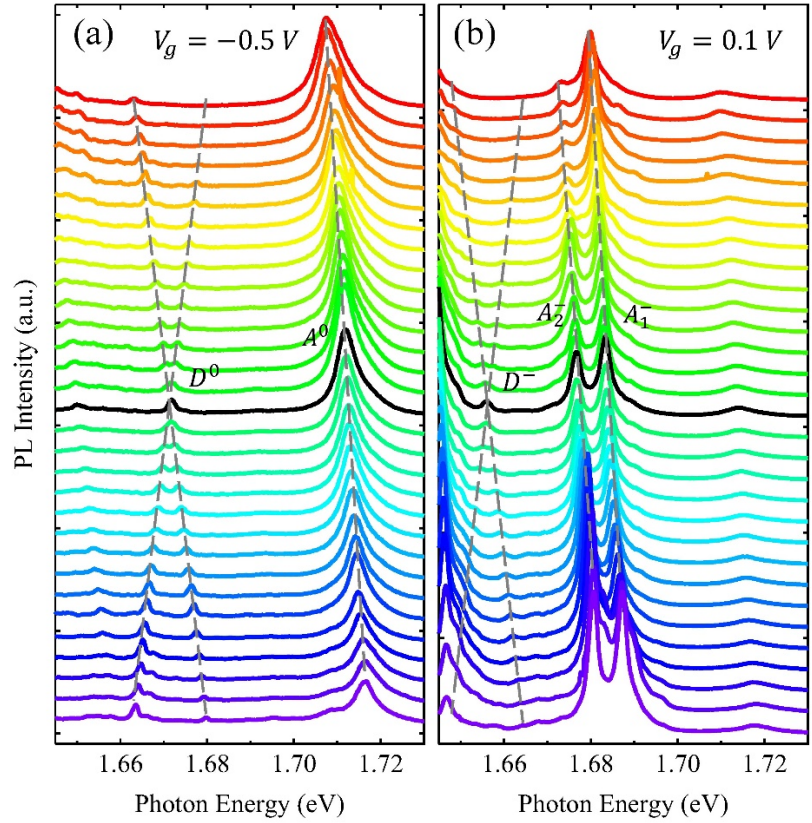

**Fig. S9.** (a-b) Magnetic-field dependent PL spectra of monolayer WSe$_2$ at gate voltages $V_g$ = –0.5 V and 0.1 V. The magnetic field increases from -30 to +30 T from bottom to top with 2 T even increment. The black lines are the spectra at zero magnetic field. The dashed lines highlight the valley Zeeman shifts. These spectra correspond to the PL color map in Fig. 3(c) of the main paper.

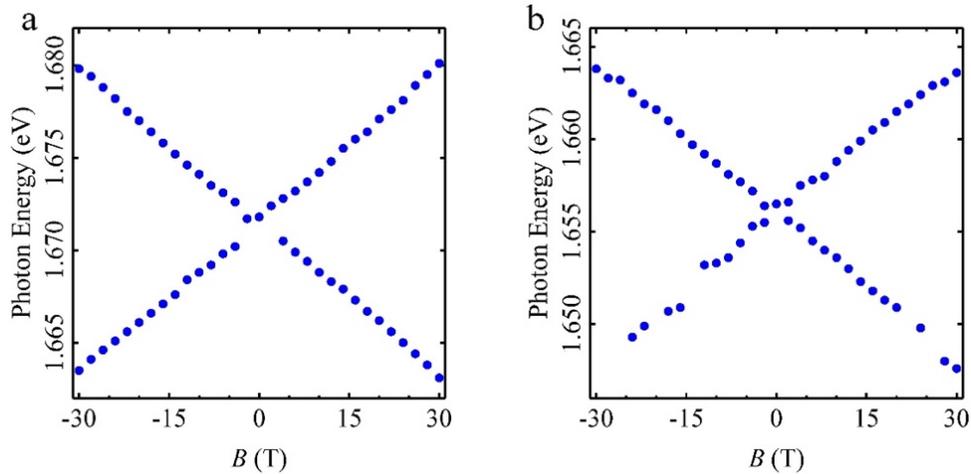

**Fig. S10.** Valley Zeeman splitting of (a) dark exciton ($D^0$) and (b) dark trion ($D^-$). The energies are extracted from Fig. S9 and Fig. 3c in the main paper.

## 10. Supplementary results and analysis of time-resolved photoluminescence

In Fig. 4 of the main paper, we present the time-resolved PL traces of the positive dark trion ($D^+$) only in two gate voltages ($V_g$ = –0.8 and –3 V). Below we show the time-resolved trion PL traces at other gate voltages (Fig. S11a) as well as the time-resolved PL traces of the bright and dark excitons at the charge neutrality point (Fig. S11b).

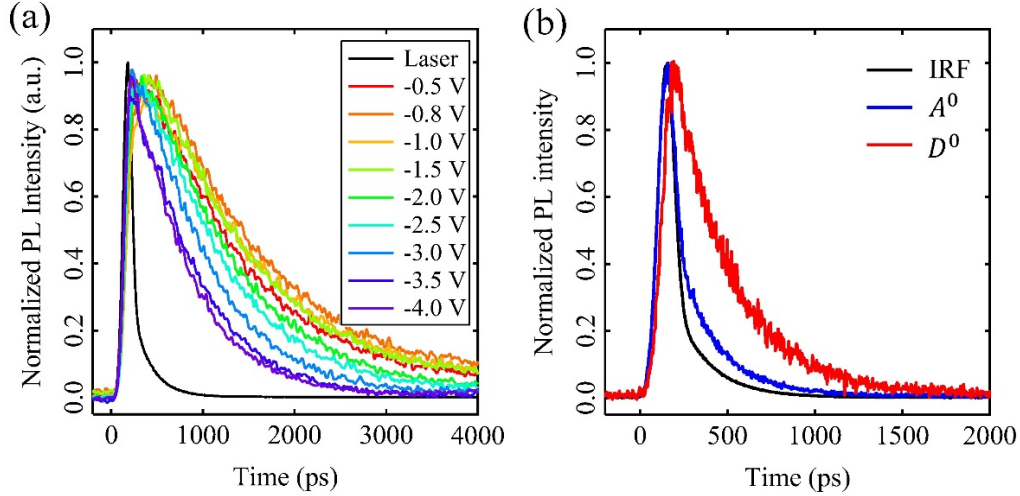

**Fig. S11.** (a) Time-resolved PL of positive dark trions ($D^+$) at different gate voltages ($V_g$). The decay process slows down slightly from $V_g$ = –0.5 to –0.8 V and becomes faster from $V_g$ = –0.8 to –4.0 V. The black line is the intrusment response function (IRF), taken directly from the laser pulses. (b) Time-resolved PL of dark exciton ($D^0$) and bright exciton ($A^0$). We extract a lifetime of 180 ps for the dark exciton and a lifetime shorter than 4 ps after deconvolution with the IRF.

We have used exponential functions convolved with the instrument response function (IRF) to fit the time-resolved PL traces. In particular, we have described the bright trion dynamics by two expoential decay processes. Fig. S12 shows our fitting of the bright trion ($A^+$) dynamics by an biexpoential function $I(t) = A_1 e^{-t/\tau_1} + A_2 e^{-t/\tau_2}$ in convolution with the IRF. We get the best fit parameters as $A_1 = 0.95, \tau_1 = 10\ ps$ for the fast component and $A_2 = 0.035, \tau_2 = 252\ ps$ for the slow component. The fast component has a much larger magnitude than the slow component and hence dominates the dynamics at the short time scale. But the weak slow component can still produce a noticeacible long tail after convolution with the IRF.

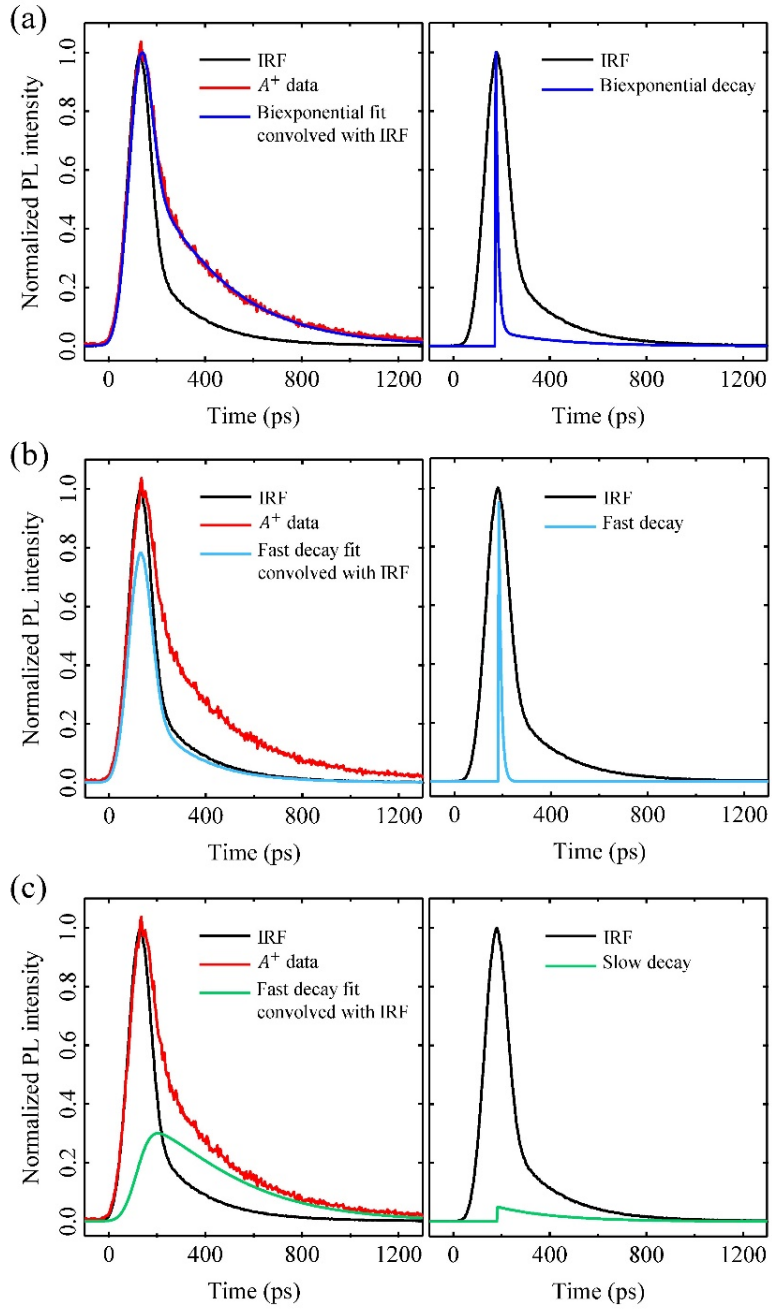

**Fig. S12.** Fitting of the time-resolved PL trace of positive bright trion (A$^+$). (a) The left panel shows the biexponential fit convolved with the instrument response function (IRF). The right panel shows the biexponential fit without convolution. (b) The left panel shows the fast component of the biexponential fit convolved with the IRF. The right panel shows the fast decay without convolution. (c) The left panel shows the slow component of the biexpoential fit convolved with the IRF. The right panel shows the slow decay without convolution.

**Supplemental References:**